\documentstyle[aps,twocolumn]{revtex}

\begin{document}

\title{Final state interactions in two-particle interferometry}

\author{D.~Anchishkin$^{a,b,}$\cite{email1}, U.~Heinz$^{a,}$\cite{email2},
and P.~Renk$^{a,}$\cite{email3}}

\address{$^a$Institut f\"ur Theoretische Physik, Universit\"at Regensburg,\\
       D-93053 Regensburg, Germany}
\address{$^b$Bogolyubov Institute for Theoretical Physics, \\
National Academy of Sciences of Ukraine, 252143 Kiev-143, Ukraine}

\date{\today}

\maketitle

\begin{abstract}

We reconsider the influence of two-particle final state interactions
(FSI) on two-particle Bose-Einstein interferometry. We concentrate in
particular on the problem of particle emission at different times.
Assuming chaoticity of the source, we derive a new general expression
for the symmetrized two-particle cross section. We discuss the
approximations needed to derive from the general result the
Koonin-Pratt formula. Introducing a less stringent version of the
so-called smoothness approximation we also derive a more accurate
formula. It can be implemented into classical event generators and
allows to calculate FSI corrected two-particle correlation functions 
via modified Bose-Einstein "weights".

\end{abstract}

\pacs{PACS numbers: ...................................}

%%%%%%%%%%%%%%%%%%%%%%%%%%%%%%%%%%%%%%%%%%%%%%%%%%%%%%%%%%%%%%%%%%%%%
\section{Introduction}
\label{sec1}
%%%%%%%%%%%%%%%%%%%%%%%%%%%%%%%%%%%%%%%%%%%%%%%%%%%%%%%%%%%%%%%%%%%%%

Two-particle correlations in momentum space can be used to extract
information about the space-time structure of the emitting source
\cite{boal90,heinz97}. The method exploits in an essential way the
quantum mechanical uncertainty relation between coordinates and
momenta, and thus any formal treatment of two-particle correlations
must be based on a quantum mechanical description. For so-called
"chaotic" sources where the two particles are emitted independently
the description can be based on the {\it single-particle} Wigner
density $S(x,K)$ of the source. It is, however, known to be important
(at least in principle, but for sufficiently small sources also in
practice) that one starts from the correct quantum mechanical
Wigner density rather than directly from a classical space
distribution $S_{\rm class}(x,K)$ of the particles in the source
because the latter can lead to unphysical behavior of the correlation
function \cite{zhang97}. The correlations are ``generated'' after the
emission by two classes of effects: 

(1) For pairs of identical bosons (fermions) the two-particle wave
function describing their propagation towards the detector must be
symmetrized (antisym\-me\-trized). For boson pairs this results in quantum
statistical ``Bose-Einstein correlations'' between the final state
momenta of the two particles. Via the uncertainty relation these
momentum space correlations reflect the spatial and temporal structure
of the source from which the two particles were emitted. This
correspondence forms the basis for Bose-Einstein interferometry in
nuclear and particle physics \cite{goldhaber60}, a variant of the
well-known Hanbury Brown--Twiss intensity interferometry in
astrophysics \cite{hanbury54}. 

(2) For both pairs of identical and non-identical particles additional
two-particle correlations can be generated by two-particle final state
interactions, notably by the long-range Coulomb interaction if the
particles carry charge \cite{gkw,pratt86,pratt90,bowler88,baym96,pratt95}.
These final state effects also depend on the spatial and temporal
distance between the emission points of the two particles and thus 
must contain information about the space-time structure of the
source. While in the last few years we have seen considerable progress
in our understanding of how to extract from two-particle correlations
quantitative information on both the geometric and the dynamic
space-time structure (sizes, lifetime and expansion velocities) of the
source {\em in the absence of final state interactions (FSI)}
\cite{heinz97,heinz96}, not much is known about how to 
{\em quantitatively} correct these methods for FSI effects. This is 
largely due to the lack of a general and exact expression which
relates the measured correlation function to the emission function of
the source. In the absence of FSI (and for chaotic source) this
relation reads \cite{shuryak73,pratt84,chapman94,gkw}
 \begin{equation}
   C({\bf q,K}) = 1 \pm
   \frac{ \left\vert \int_x e^{i q\cdot x }\, S(x,K)\right\vert^2 }
        { \int_x S\left(x, K{+}{\textstyle{q\over 2}}\right)  
          \int_y S\left(y, K{-}{\textstyle{q\over 2}}\right) } \, ,
 \label{1}
 \end{equation}
where $S(x,K)$ is the single particle Wigner density (``emission
function'') of the source, and where 
 \begin{eqnarray}
 \label{1a}
  {\bf K} &=& {\textstyle{1\over 2}}({\bf p}_a + {\bf  p}_b)\, , \quad
  K^0={\textstyle{1\over 2}}(E_a+E_b)\, ,
 \nonumber\\
  {\bf q} &=& {\bf p}_a - {\bf p}_b\, ,\qquad
  q^0 = E_a - E_b = {\bf q} \cdot {\bf K}/K^0\, ,
 \end{eqnarray}
with $E_{a,b}=\sqrt{m^2+{\bf p}_{a,b}^2 }$. The second term in
(\ref{1}) reflects the quantum statistical correlations. This
expression allows to expand the correlation function $C({\bf q,K})$ in
space-time moments of the emission function \cite{heinz96,tomasik96}
and relate the main characteristics of the two-particle correlator,
its width as a function of the relative momentum ${\bf q}$, to the
second central space-time moment (``r.m.s. widths'') of the emission 
function in coordinate space. The factor $e^{i q{\cdot}x}$ in the
correlation term in (\ref{1}) reflects the assumed absence of FSI, i.e.
plane wave propagation.

The goal of the present paper is to generalize relation (\ref{1}) to
the case including two-particle FSI. Existing treatments of this problem
\cite{pratt86,pratt90,bowler88,pratt95,lednicky} differ in their results
even on the formal level because different types of approximations are
used already in intermediate stages of the calculation. This,
unfortunately, makes a direct comparison of these results and a check
of the approximations essentially impossible. Our aim is to give a
rigorous derivation of the generalization of Eq.~(\ref{1}), using only
the following two approximations: 

(1) the source is completely chaotic, i.e. the particles are emitted
independently, and

(2) finite multiplicity corrections \cite{gkw,wiedemann97} can be
neglected.\newline 
Both approximations are expected to be good for high energy nuclear
collisions with large multiplicities, but may require further scrutiny
in lower multiplicity $e^+ e^-$ or hadron-hadron collisions. On the
other hand, high multiplicities can result in modifications
(screening) of the two-particle (Coulomb) potential due to the
influence of the environment of other charged particles. It was,
however, shown in \cite{anchishkin96} that such screening effects are
small and can be neglected because the charge density of the
environment drops rapidly as a function of the distance from the
collision center (faster than $1/r^2$) and the pair leaves very
quickly the region of high charge density.

%%%%%%%%%%%%%%%%%%%%%%%%%%%%%%%%%%%%%%%%%%%%%%%%%%%%%%%%%%%%%%%%%%%%%%%%%%%
\section{The two-particle cross section}
\label{sec2}
%%%%%%%%%%%%%%%%%%%%%%%%%%%%%%%%%%%%%%%%%%%%%%%%%%%%%%%%%%%%%%%%%%%%%%%%%%%
\subsection{General strategy}
\label{sec2a}
%%%%%%%%%%%%%%%%%%%%%%%%%%%%%%%%%%%%%%%%%%%%%%%%%%%%%%%%%%%%%%%%%%%%%%%%%%%

The two-particle correlation function $C({\bf p}_a,{\bf p}_b)
\equiv C({\bf q,K})$ is defined as a ratio between the
normalized invariant two-particle coincidence cross section
$P_2({\bf p}_a,{\bf p}_b)$ and the product of the invariant single
particle cross sections $P_1({\bf p}_{a,b})$:
 \begin{equation}
  C({\bf q,K}) =
  \frac{P_2({\bf p}_a,{\bf p}_b)}
       {P_1({\bf p}_a)\, P_1({\bf p}_b)} \, ,
\label{2}
\end{equation}
where $P_2({\bf p}_a,{\bf p}_b) = E_a E_b\frac{d^6N}{d^3p_a d^3p_b}$,
$P_1({\bf p}_{a,b}) = E_{a,b} \frac{d^3N}{d^3p_{a,b}}$. Most space
will be taken by the calculation of the two-particle emission
probability $P_2({\bf p}_a,{\bf p}_b)$. Its normalization by the
product of single particle probabilities is a trivial final step.

Our strategy is as follows: We start by writing down an expression for
the two-particle probability amplitude
$A_\gamma({\bf p}_a,{\bf p}_b)$ for measuring at $t \to \infty$ 
in the detector a pair of particles with momenta ${\bf p}_a$ and
${\bf p}_b$ if the pair has been emitted by the source in a
two-particle state $\psi_\gamma$. The square of this amplitude,
averaged properly over the distribution of two-particle quantum
numbers $\gamma$ in the source by taking the trace with an appropriate
density matrix for the source, will give the two-particle probability
$P_2({\bf p}_a,{\bf p}_b )$. This amplitude, first written
down in terms of an overlap integral at $t \to \infty$ with an
asymptotic two-particle momentum eigenstate in the detector, is then
rewritten in terms of a corresponding overlap integral in the source 
at the time of freeze-out, by using the time evolution operator
(including the two-particle FSI) to evolve the asymptotic state
backwards in time.

Here a crucial ingredient will be the realization that the two-particle
FSI can only act while both particles are present. If one is emitted
earlier than the other, in the absence of one-body final state
interactions with the remainder of the source (e.g. between its charge
and that of the remaining fireball which we will here neglect) it will
undergo {\it free} time evolution until the second particle has also
been formed. To implement this requires the factorization of the
two-particle wave function into a product of single-particle states,
i.e. the assumption of independent emission. This interval of free
propagation gives rise to an additional phase which contains the time
structure of the source and which may be important for attempts to
extract source lifetimes from the two-particle correlator. As far as
we know, this aspect has not been fully discussed in previous FSI 
studies \cite{pratt86,pratt90,bowler88,baym96,pratt95,lednicky}.

Once the two-particle amplitude $A_\gamma({\bf p}_a,{\bf p}_b)$
is expressed in terms of an overlap integral at the time of particle
freeze-out, one can calculate the two-particle probability by
averaging over the quantum numbers $\gamma$ and the distribution of
emission times. Due to the assumed factorization of the two-particle
wave function the result can be expressed in terms of the single
particle Wigner density of the source at freeze-out $S(x,K)$ which
will be defined in Eq.~(\ref{42}). The remainder of the derivation
consists of an appropriate rewriting of this expression which allows
to compare it with Eq.~(\ref{1}) and which can serve as a starting
point for further approximations in order to compare with previously
published expressions in Refs.~\cite{pratt86,pratt90,pratt95}.

We should stress that our derivation is intrinsically non-relativistic.
For this to be appropriate we must work in a reference frame in which
the pair moves non-relativistically, i.e. where ${\bf K}\approx 0$.
Since we are interested in the deviation of the correlator
$C({\bf q,K})$ from unity for small ${\bf q}$, the relative
motion is then also non-relativistic. In this frame the two-particle
FSI can be represented by an instantaneous potential, and the
difference between the emission times of the two particles is
well-defined. The final result can then be evaluated in an arbitrary
frame by proper Lorentz transformation of all momenta. A completely
covariant derivation (which should also be possible by introducing
propagating fields for the FSI) has not yet been achieved. 

%%%%%%%%%%%%%%%%%%%%%%%%%%%%%%%%%%%%%%%%%%%%%%%%%%%%%%%%%%%%%%%%%%%%%%%
\subsection{The two-particle momentum amplitude}
\label{sec2b}
%%%%%%%%%%%%%%%%%%%%%%%%%%%%%%%%%%%%%%%%%%%%%%%%%%%%%%%%%%%%%%%%%%%%%%%

Let us consider a two-particle state $\psi_\gamma$ emitted by the
source. Its propagation to the detector is governed by the
Schr\"odinger equation 
 \begin{equation}
  i \frac{\partial \psi_\gamma({\bf x}_a,{\bf x}_b,t)}
         {\partial t} = 
  \hat{H}({\bf x}_a,{\bf x}_b)\,
  \psi_\gamma({\bf x}_a,{\bf x}_b,t) \, ,
 \label{3}
 \end{equation}
where
 \begin{equation}
  \hat{H}({\bf x}_a,{\bf x}_b) = \hat{h} ({\bf x}_a) + \hat{h} ({\bf x}_b) +
  V\left(\left\vert {\bf x}_a-{\bf x}_b \right\vert \right) \, ,
 \label{4}
 \end{equation}
and
 \begin{equation}  
  \hat{h} ({\bf x}_i) = m - \frac{1}{2m} \bbox{\nabla}_i^2 \, .
 \label{5}
 \end{equation}
The index $\gamma$ denotes a complete set of 2-particle quantum numbers.
(In a basis of products of two wave packets these could contain the
centers ${\bf X}_a$, ${\bf X}_b$ of the wavepackets of the two
particles at their freeze-out times $t_a$, $t_b$, respectively.)
Eq.~(\ref{3}) is solved by
 \begin{equation}
  \psi_\gamma({\bf x}_a,{\bf x}_b,t) = 
  e^{- i \hat{H}({\bf x}_a,{\bf x}_b) (t-t_0)}\,
  \psi_\gamma({\bf x}_a,{\bf x}_b,t_0) 
 \label{6}
 \end{equation}
in terms of the two-particle wave function at some initial time $t_0$.
We will assume that the detector measures asymptotic momentum
eigenstates, i.e. that it acts by projecting the emitted 2-particle
state onto 
 \begin{eqnarray}
  \phi^{\rm out}_{{\bf p}_a,{\bf p}_b}({\bf x}_a,{\bf x}_b,t)
  &=& e^{i ({\bf p}_a\cdot {\bf x}_a - E_a t)}\,
    e^{i ({\bf p}_b\cdot {\bf x}_b - E_b t)} 
 \nonumber\\
  &=& e^{- i E t}\, 
    e^{i ({\bf p}_a\cdot {\bf x}_a + {\bf p}_b\cdot {\bf x}_b)} \, ,
 \label{7}
 \end{eqnarray}
where $E = E_a + E_b = 2 K^0$ with $E_{a,b} = \sqrt{m^2+{\bf p}_{a,b}^2}$
is the total energy of the pair. In this paper we will only consider
the case of pairs of identical particles, $m_a=m_b=m$;
correspondingly, the two-particle states $\psi_\gamma$ in (\ref{6})
must be (anti-)symmetrized (see below). The measured two-particle
momentum amplitude is then 
 \begin{eqnarray}
   && A_\gamma({\bf p}_a,{\bf p}_b) =
 \nonumber\\
   && \lim_{t\to\infty} 
   \int d^3x_a\, d^3x_b\, 
   \phi^{\rm out,*}_{{\bf p}_a,{\bf p}_b} ({\bf x}_a,{\bf x}_b,t) \,
   \psi_\gamma({\bf x}_{a},{\bf x}_{b},t) \, .
 \label{8}
 \end{eqnarray}
Using the time evolution equation (\ref{6}) this can be expressed in
terms of the emitted two-particle wave function $\psi_\gamma$ at
earlier times as 
 \begin{eqnarray}
  && A_\gamma({\bf p}_a,{\bf p}_b) = \lim_{t\rightarrow \infty}
   \int d^3x_a\, d^3x_b\,
   \psi_\gamma({\bf x}_a,{\bf x}_b,t_0) 
 \nonumber\\
  && \quad \times
   \left[ \exp[ - i \hat{H}({\bf x}_a,{\bf x}_b) (t_0-t)] \,
          \phi^{\rm out}_{{\bf p}_a,{\bf p}_b} ({\bf x}_a,{\bf x}_b,t)
   \right]^* \, .
 \label{9}
 \end{eqnarray}
This is correct for all times $t_0 \ge {\rm max}\, [t_a,t_b]$ where
$t_{a,b}$ are the freeze-out times for the two particles. Note that by
partial integration (or inversion of the unitary evolution operator)
the time evolution operator has been shifted from the emitted
two-particle state (with arbitrary quantum numbers $\gamma$) 
to the two-particle momentum eigenstate $\phi^{\rm out}_{{\bf p}_a,
  {\bf p}_b} ({\bf x}_a,{\bf x}_b,t)$ which thereby is transformed
into a distorted wave at time $t_0$ which includes the effect of the
FSI in such a way that after evolution from $t_0$ to $t=\infty $ it
again becomes a plane wave with momentum ${\bf p}_a$, ${\bf p}_b$ and
energy $E=E_a+E_b$. 

%%%%%%%%%%%%%%%%%%%%%%%%%%%%%%%%%%%%%%%%%%%%%%%%%%%%%%%%%%%%%%%%%%%%%%%
\subsection{The distorted wave}
\label{sec2c}
%%%%%%%%%%%%%%%%%%%%%%%%%%%%%%%%%%%%%%%%%%%%%%%%%%%%%%%%%%%%%%%%%%%%%%%

To further analyze Eq.~(\ref{9}) we introduce center-of-mass and
relative coordinates according to 
 \begin{eqnarray}
  {\bf R}&=&\frac{1}{2} ({\bf x}_a+{\bf x}_b) \, ,\quad
  {\bf P}= 2{\bf K}={\bf p}_a+{\bf p}_b \, ,
 \nonumber \\
  {\bf r}&=&{\bf x}_a-{\bf x}_b \, , \qquad
  {\bf q}={\bf p}_a-{\bf p}_b \, ,
 \label{10}
 \end{eqnarray}
such that
 \begin{equation}
   \bbox{\nabla}_a^2 + \bbox{\nabla}_b^2 = 
   \frac{1}{2} \bbox{\nabla}_R^2 + 2 \bbox{\nabla}_r^2 \, ,
 \label{11}
 \end{equation}
and
 \begin{eqnarray}
  \hat{H} ({\bf x}_{a},{\bf x}_{b}) &=&
  \hat{H}_0({\bf R}) + \hat{H}_1({\bf r}) \, ,
 \nonumber \\
  \hat{H}_0({\bf R}) &\equiv& - \frac{1}{2M} \bbox{\nabla}_R^2 \, ,
 \nonumber \\
  \hat{H}_1({\bf r}) &\equiv& - \frac{1}{2 \mu} \bbox{\nabla}_r^2 +
  V(r) \, ,
 \label{12}
 \end{eqnarray}
with $M=2m$ and $\mu =m/2$ for identical particles. In these
coordinates the asymptotic wave function (\ref{7}) reads 
 \begin{equation}
   \phi^{\rm out}_{{\bf p}_a,{\bf p}_b} ({\bf x}_a,{\bf x}_b,t) =
   e^{i ({\bf P}\cdot {\bf R} - Et)} \, 
   e^{\frac{i}{2} {\bf q}\cdot {\bf r}} \, . 
 \label{13}
 \end{equation}
In Eq.~(\ref{9}) we must evaluate the action of the time evolution
operator on this function. To this end let us separate the total
energy $E=2K^0$ into its contributions from the c.m. and relative
motions, $E=E_{\rm cm}+E_{\rm rel}$, where $E_{\rm cm} = \sqrt{M^2 +
  {\bf P}^2}=2\sqrt{m^2+{\bf K}^2} \equiv 2E_K \approx M+{\bf P}^2/2M$
and $E_{\rm rel} = E - E_{\rm cm} = 2 (K^0-E_K) \approx {\bf q}^2/(4E_K)
\approx {\bf q}^2/4m$ in a frame where ${\bf K} \approx 0$. The action
of $\hat{H}_0({\bf R})$ on the plane wave factor for the c.m. motion
is then easily evaluated: 
 \begin{equation}
  e^{ - i \hat{H} _0({\bf R}) (t_{0}-t)}\,  e^{i {\bf P}\cdot {\bf R}}
  = e^{ - i \sqrt{M^2+{\bf P}^2}\, (t_{0}-t)} \,
  e^{i {\bf P}\cdot {\bf R}} \, .
 \label{14}
 \end{equation}
What is left is the time evolution of the wave function for the
relative motion:
 \begin{equation}
  \lim_{t \to \infty }\, e^{-i \hat{H}_1({\bf r}) (t_0-t)}
  \, e^{\frac{i}{2} {\bf q}\cdot {\bf r}} \, e^{-i E_{\rm rel} t} \equiv
  e^{-i E_{\rm rel} \, t_0} \, \phi_{{\bf q}/2}({\bf r}) \, .
 \label{15}
 \end{equation}
Here $\phi_{{\bf q}/2}({\bf r})$ is a solution of the stationary
Schr\"odinger equation 
 \begin{equation}
   \hat{H}_1({\bf r}) \, \phi_{{\bf q}/2}({\bf r}) = E_{\rm rel}
   \phi_{{\bf q}/2}({\bf r})  \, ,
 \label{16}
 \end{equation}
(where $E_{\rm rel} = \frac{{\bf q}^2}{4E_K}$) with asymptotic
boundary conditions 
 \begin{equation}
  \lim_{|{\bf r}| \to \infty} \phi_{{\bf q}/2}({\bf r}) =
  e^{\frac{i}{2} {\bf q}\cdot {\bf r}} \, .
 \label{17}
 \end{equation}
Inserting everything into Eq.~(\ref{9}) we find
 \begin{eqnarray}
 \label{19}
   A_\gamma({\bf p}_a,{\bf p}_b) &=& 
   e^{i E t_0} \int d^3r\, d^3R \, e^{-i {\bf P}\cdot {\bf R} }
   \phi^*_{{\bf q}/2}({\bf r})
 \nonumber\\
   && \times \
   \psi_\gamma\left({\bf R}+{\textstyle{{\bf r}\over 2}},
                    {\bf R}-{\textstyle{{\bf r}\over 2}},t_0\right) \, .
 \end{eqnarray}
%

%%%%%%%%%%%%%%%%%%%%%%%%%%%%%%%%%%%%%%%%%%%%%%%%%%%%%%%%%%%%%%%%%%%%%%%%%%%%
\subsection{Free propagation between emission points}
\label{sec2d}
%%%%%%%%%%%%%%%%%%%%%%%%%%%%%%%%%%%%%%%%%%%%%%%%%%%%%%%%%%%%%%%%%%%%%%%%%%%%

With Eq.~(\ref{19}) we have rewritten the measured two-particle
momentum space amplitude $A_\gamma({\bf p}_a,{\bf p}_b)$ in terms of
an overlap integral evaluated at the earliest time $t_0$ where both
particles are present. Let us, for the moment, call this time
$t_a$. At this time the first emitted particle of the pair has already
propagated for a time $t_a-t_b$ if it was emitted at $t_b <
t_a$. During this time the first emitted particle cannot ``see'' the
second particle as a separate entity, but only as part of the remaining
fireball. Nevertheless, its charge is there, ``hidden'' among the
charges of all the other particles in the fireball.

This stage of the time evolution is obviously very complicated.
The idealization of letting the two particles interact via a two-body
(Coulomb) potential clearly breaks down, since the second particle has
not been ``formed'' yet. On the other hand, considering the first
emitted particle as freely propagating during this time is not
necessarily a good approximation either: there should at least be the
interaction with charge of the rest of the fireball which during this
time interval is one unit larger than after emission of the second
particle. However, considering both the interaction between the two
particles and between each one of them and the remaining fireball is a
non-trivial quantum mechanical tree-body problem, at least until the
pair has well separated from the fireball.

We will therefore study here only the simple approximation that the
first emitted particle propagates freely until the second particle
freezes out. In order to implement this into the formalism we must
make the assumption that the two particles are emitted independently,
i.e. that at $t_a$ (up to symmetrization effects) the two-particle
wave function $\psi_\gamma$ factorizes:
 \begin{eqnarray}
   \psi_\gamma({\bf x}_a,{\bf x}_b,t_a) = \frac{1}{\sqrt{2}}
   &&\left[ \psi_{\gamma_a}({\bf x}_a,t_a)\, 
          \psi_{\gamma_b}({\bf x}_b,t_a) \right.
 \nonumber\\
   &&\pm \left. 
          \psi_{\gamma_a}({\bf x}_b,t_a)\, 
          \psi_{\gamma_b} ({\bf x}_a,t_a) \right] \, .
 \label{20}
 \end{eqnarray}
The indices $\gamma_a,\gamma_b$ on the 1-particle wave functions now
label complete sets of 1-particle quantum numbers. For simplicity of
notation we will from now on replace $\gamma_{a,b}$ by $a,b$.  

Let us first consider the case that the particle in the state $\psi_b$
was emitted earlier, at time $t_b < t_a$. We can then write (see
Eq.~(\ref{4})) 
 \begin{equation}
  \psi_b({\bf x},t_a) = e^{-i \hat{h} ({\bf x}) (t_a-t_b)}
  \psi_b ({\bf x},t_b) \, ,
 \label{21}
 \end{equation}
and hence
 \begin{eqnarray}
   \psi_\gamma({\bf x}_a,{\bf x}_b,t_a) &=&
 \label{22}\\
   {\theta(t_a-t_b)\over\sqrt{2}} \!\!&&\!\!\left[
      \psi_a({\bf x}_a,t_a)\, e^{-i \hat{h} ({\bf x}_b) (t_a-t_b)}
      \psi_b({\bf x}_b,t_b) \right.
 \nonumber\\
  &&\!\!\pm \left.
      \psi_a({\bf x}_b,t_a)\, e^{-i \hat{h} ({\bf x}_a) (t_a-t_b)}
      \psi_b({\bf x}_a,t_b) \right] \, .
 \nonumber
 \end{eqnarray}
In order to evaluate the action of the free time-evolution operator we
Fourier decompose $\psi_a$, $\psi_b$ into momentum eigenstates:
 \begin{mathletters}
 \label{23}
 \begin{eqnarray} 
  \psi_a({\bf x},t) &=& \int \frac{d^3 k_a}{(2\pi )^3}\,
  \tilde{\psi}_a({\bf k}_a,t) \, 
  e^{i ({\bf k}_a\cdot {\bf x}- \omega _a t)} \, , 
 \label{23a}\\
 \omega_a &=& \sqrt{m^2+{\bf k}_a^2} \, ,
 \label{23b}
 \end{eqnarray}
 \end{mathletters}
and similarly for $\psi_b$ with integration variable ${\bf k}_b$.
One then obtains
 \begin{eqnarray}
  && \psi_\gamma({\bf x}_a,{\bf x}_b,t_a) = {\theta(t_a-t_b)\over\sqrt{2}} 
  \int \frac{d^3 k_a}{(2\pi )^3}\, \frac{d^3 k_b}{(2\pi )^3}\,
 \nonumber\\
  && \quad \times \, 
  \tilde{\psi }_a ({\bf k}_a,t_a)\, \tilde{\psi }_b ({\bf k}_b,t_b) \
   e^{-i [ \omega _a t_a+\omega _b(t_a-t_b)+\omega _bt_b]} 
 \nonumber\\
  && \quad \times \, 
  \left[
   e^{i ( {\bf k}_a\cdot {\bf x}_a+{\bf k}_b\cdot {\bf x}_b)} \pm
   e^{i ( {\bf k}_b\cdot {\bf x}_a+{\bf k}_a\cdot {\bf x}_b)}
  \right] \, ,
 \label{24}
 \end{eqnarray}
where we have written out in the time factor separately the contributions
from the Fourier transform and from the free time evolution according to
(\ref{22}). Note that the terms $\sim t_b$ cancel between them.

Inserting (\ref{24}) into (\ref{19}) and defining
 \begin{equation}
  {\bf Q}= {\bf k}_a+{\bf k}_b \, , \quad
  {\bf k}=\frac{1}{2} ({\bf k}_a-{\bf k}_b) 
 \label{25}
 \end{equation}
we arrive at
 \begin{eqnarray}
  && A_{ab}^{(a)}({\bf p}_a,{\bf p}_b) = {\theta(t_a-t_b)\over\sqrt{2}}
  \int \frac{d^3 Q}{(2\pi )^3} \, \frac{d^3 k}{(2\pi )^3} \,
       d^3r \, d^3R\,
 \nonumber\\
  &&\quad \times\ 
  e^{i ({\bf Q}-{\bf P})\cdot {\bf R}} \,
  e^{i {\bf k}\cdot {\bf r}}\, 
  e^{i \Omega ({\bf Q}, {\bf k})t_a } \left[
     \phi^*_{ {\bf q}/2}({\bf r}) \pm \phi^*_{ -{\bf q}/2}({\bf r})
     \right]
 \nonumber\\
  && \quad \times\  
     \tilde{\psi}_a \left({\textstyle{{\bf Q}\over 2}}+{\bf k},t_a\right) \,
     \tilde{\psi}_b \left({\textstyle{{\bf Q}\over 2}}-{\bf k},t_b\right) \, .
 \label{26}
 \end{eqnarray}
The superscript $(a)$ in the amplitude should remind us that the
appearance of the second particle (i.e. the creation of the ``pair'')
happened at time $t_a$. In deriving (\ref{26}) we used the fact that
both $\phi_{{\bf q}/2}({\bf r})$ and $\phi_{-{\bf q}/2}({\bf r})$ are
solutions to (\ref{16}) with the same energy eigenvalue, but
corresponding to boundary conditions (\ref{17}) with opposite signs in
the exponent. From this it follows immediately that 
 \begin{equation}
  \phi_{-{\bf q}/2}({\bf r}) = \phi_{{\bf q}/2}(-{\bf r}) \, .
 \label{18}
 \end{equation}
With this trick the symmetrization effects on the emitted two-particle
state can be absorbed into the relative wavefunction describing the
FSI. -- In (\ref{26}) we also replaced the two-particle quantum number
index $\gamma$ by the pair $(ab)$ of single particle quantum numbers
and defined  
 \begin{equation}
  \Omega \left({\bf Q},{\bf k}\right) \equiv 
  E - \omega \left({\textstyle{{\bf Q}\over 2}}+{\bf k}\right) 
    - \omega \left({\textstyle{{\bf Q}\over 2}}-{\bf k}\right) \, .
 \label{27}
 \end{equation}

The integration over the c.m. coordinate ${\bf R}$ of the pair is
trivial and yields $(2\pi)^3 \,\delta ^3({\bf Q}-{\bf P})$. After
doing the integrations over ${\bf Q}$ and ${\bf r}$ we get (remember
that ${\bf P}= 2{\bf K}={\bf p}_a+{\bf p}_b$) 
 \begin{eqnarray}
  A_{ab}^{(a)}({\bf p}_a,{\bf p}_b) &=& \theta(t_a-t_b)
  \int \frac{d^3 k}{(2\pi )^3} \, e^{i \Omega({\bf K},{\bf k})t_a }\,
  \tilde{\Phi}^*\left({\textstyle{{\bf q}\over 2}},{\bf k}\right) \,
 \nonumber\\
 &&\quad\times\ 
  \tilde{\psi}_a({\bf K}+{\bf k},t_a)\, \tilde{\psi}_b({\bf K}-{\bf k},t_b)
  \, .
 \label{28}
 \end{eqnarray}
Here
 \begin{equation}
   \Omega({\bf K,k}) \equiv E_a - \omega({\bf K}+{\bf k}) 
                          + E_b - \omega({\bf K}-{\bf k}) \, ,
 \label{29}
 \end{equation}
and we defined the symmetrized FSI distorted wave
 \begin{equation}
   \Phi\left({\textstyle{{\bf q}\over 2}},{\bf k}\right) = {1\over\sqrt{2}}
   \left[ \tilde{\phi}\left({\textstyle{{\bf q}\over 2}},{\bf k}\right) \pm
          \tilde{\phi}\left(-{\textstyle{{\bf q}\over 2}},{\bf k}\right)
 \right]\, , 
 \label{36}
 \end{equation}
where
 \begin{equation}
   \tilde{\phi}\left({\textstyle{{\bf q}\over 2}},{\bf k}\right) =
   \int d^3r\, e^{-i {\bf k}\cdot {\bf r}} \, \phi_{{\bf q}/2}({\bf r})
 \label{30}
 \end{equation}
is the momentum space representation of the distorted relative
wave function with asymptotic relative momentum ${\bf q}/2$.

The opposite situation that the particle in the state $\psi_a$ was
emitted earlier at time $t_a$ and the ``pair'' appears only later at
time $t_b$ can be dealt with in a similar way. One finds
 \begin{eqnarray}
  A_{ab}^{(b)}({\bf p}_a,{\bf p}_b) &=& \theta(t_b-t_a)
  \int \frac{d^3 k}{(2\pi )^3} \, e^{i \Omega({\bf K},{\bf k})t_b }\,
  \Phi^*\left({\textstyle{{\bf q}\over 2}},{\bf k}\right) \,
 \nonumber\\
  && \quad \times\
  \tilde{\psi}_a({\bf K}+{\bf k},t_a)\, \tilde{\psi}_b({\bf K}-{\bf k},t_b)
  \, .
 \label{28a}
 \end{eqnarray}
The total amplitude is the sum of the contributions from the two
different time orderings:
 \begin{equation}
 \label{sum}
   A_{ab}({\bf p}_a,{\bf p}_b) = A_{ab}^{(a)}({\bf p}_a,{\bf p}_b)
                               + A_{ab}^{(b)}({\bf p}_a,{\bf p}_b)\, .
 \end{equation}

Let us note for later reference that neglecting the free time
evolution of the first emitted particle from time $t_b$ to time $t_a$
would have resulted in the ansatz 
 \begin{eqnarray}
  && \psi_\gamma({\bf x}_a,{\bf x}_b,t_a) =
     {\theta(t_a-t_b)\over\sqrt{2}} 
 \nonumber\\
  &&\quad\times\Bigl[\psi_a({\bf x}_a,t_a)\,\psi_b({\bf x}_b,t_b) 
    \pm \psi_a({\bf x}_b,t_a)\,\psi_b({\bf x}_a,t_b) \Bigr] 
 \label{31}
 \end{eqnarray}
instead of (\ref{22}) for the time ordering $(a)$, and similarly with
$\theta(t_a-t_b)$ replaced by $\theta(t_b-t_a)$ for the time ordering
$(b)$. This would have resulted in the modified amplitudes 
 \begin{mathletters}
 \label{32}
 \begin{eqnarray}
   A_{ab}^{(a)}({\bf p}_a,{\bf p}_b) &=& \theta(t_a-t_b)
   \,e^{i E t_a} \int \frac{d^3 k}{(2\pi )^3} \,
   \Phi^*\left({\textstyle{{\bf q}\over 2}},{\bf k}\right) \,
 \nonumber\\ &\times&
   e^{-i\omega({\bf K}+{\bf k}) t_a}\,
   \tilde{\psi}_a({\bf K}+{\bf k},t_a) 
 \nonumber\\ &\times&
   e^{-i\omega({\bf K}-{\bf k}) t_b}\,
   \tilde{\psi}_b({\bf K}-{\bf k},t_b) \, ,
 \label{32a}
 \\
   A_{ab}^{(b)}({\bf p}_a,{\bf p}_b) &=& \theta(t_b-t_a)
   \,e^{i E t_b} \int \frac{d^3 k}{(2\pi )^3} \,
   \Phi^*\left({\textstyle{{\bf q}\over 2}},{\bf k}\right) \,
 \nonumber\\ &\times&
   e^{-i\omega({\bf K}+{\bf k}) t_a}\,
   \tilde{\psi}_a({\bf K}+{\bf k},t_a) \,
 \nonumber\\ &\times&
   e^{-i\omega({\bf K}-{\bf k}) t_b}\,
   \tilde{\psi}_b({\bf K}-{\bf k},t_b) \, .
 \label{32b}
 \end{eqnarray}
 \end{mathletters}
%

%%%%%%%%%%%%%%%%%%%%%%%%%%%%%%%%%%%%%%%%%%%%%%%%%%%%%%%%%%%%%%%%%%%%%%%%
\subsection{The two-particle cross-section}
\label{sec2e}
%%%%%%%%%%%%%%%%%%%%%%%%%%%%%%%%%%%%%%%%%%%%%%%%%%%%%%%%%%%%%%%%%%%%%%%%

The two-particle cross section is obtained by averaging (\ref{sum})
and its complex conjugate with the density matrix defining the source.
This density matrix is characterized by a probability distribution for
the two-particle quantum numbers $(a,b)$ and by a distribution of
emission times $(t_a,\, t_b)$. We write
 \begin{equation}
   P_2({\bf p}_a,{\bf p}_b) =
   \sum  _{ab,a'b'} \! \! \! \! \! \! \! \! \! \! {\displaystyle  \int }
   \ \rho _{ab,a'b'}\,
   A^*_{a'b'} \left( {\bf p}_a,{\bf p}_b \right)\,
   A_{ab} \left( {\bf p}_a,{\bf p}_b \right) \, ,
 \label{33}
 \end{equation}
and make the ansatz
 \begin{equation}
   \rho _{ab,a'b'} =
   \nu _{aa'}\, \rho (t_a,t_{a'})\
   \nu _{bb'}\, \rho (t_b,t_{b'})\, .
 \label{34}
 \end{equation}
This ansatz factorizes in such a way that independent emission of the
two particles is ensured. The summation/integration in (\ref{33}) is
to be understood as 
 \begin{equation}
   \sum  _{ab,a'b'} \! \! \! \! \! \! \! \! \! \! {\displaystyle  \int }
   = \sum _{ab,a'b'} \int dt_a\, dt_b\, dt_{a'}\, dt_{b'}\,.
 \label{35}
 \end{equation}
According to (\ref{sum}) the probability consists of four terms which
we write as
 \begin{equation}
   P_2({\bf p}_a,{\bf p}_b) = 
   P^{(aa)} + P^{(bb)} + P^{(ab)} + P^{(ba)} \, .
\label{33a}
\end{equation}
We shall calculate only the first term $P^{(aa)}$ explicitly.
The second term is easily shown to equal the first one while the last
two terms will be shown to vanish.

Inserting (\ref{28}) into (\ref{33}) yields
 \begin{eqnarray}
   && P^{(aa)}({\bf p}_a,{\bf p}_b) =
   \int \frac{d^3 k}{(2\pi )^3} \, \frac{d^3 k'}{(2\pi )^3} \,
   \Phi\left({\textstyle{{\bf q}\over 2}},{\bf k}'\right) 
   \Phi^*\left({\textstyle{{\bf q}\over 2}},{\bf k}\right) 
 \nonumber\\
   &&\quad\times\,
   \int dt_a dt_b dt_{a'} dt_{b'}\, 
   \theta (t_a-t_b)\, \theta (t_{a'}-t_{b'})\,
 \nonumber\\
   &&\qquad\qquad\times\,
   e^{i(\Omega({\bf K}, {\bf k})t_a - \Omega({\bf K}, {\bf k}')t_{a'})}
 \nonumber\\
   &&\quad\times\,
   \sum_{aa'} \nu_{aa'} \, \rho (t_a,t_{a'}) \,
   \tilde{\psi}_a({\bf K}+{\bf k},t_a) \, 
   \tilde{\psi}_{a'}^*({\bf K}+{\bf k}',t_{a'}) \,
 \nonumber\\
   &&\quad\times\,
   \sum_{bb'} \nu_{bb'} \, \rho (t_b,t_{b'}) \,
   \tilde{\psi}_b({\bf K}-{\bf k},t_b) \,
   \tilde{\psi}_{b'}^*({\bf K}-{\bf k}',t_{b'}) .
 \label{37} 
 \end{eqnarray}
Let us introduce new time variables
 \begin{eqnarray}
   && X^0=\frac{1}{2} (t_a+t_{a'}) \ , \ \ \ \ \ \ x^0=t_a-t_{a'} \ , 
 \nonumber \\
   && Y^0=\frac{1}{2} (t_b+t_{b'}) \ , \ \ \ \ \ \ y^0=t_b-t_{b'} \ ,
 \label{38}
 \end{eqnarray}
as well as the inverse Fourier representation (see Eq.~(\ref{23}))
 \begin{mathletters}
 \label{39}
 \begin{eqnarray}
 &&
 \tilde{\psi}_a \left({\bf K}+{\bf k},X^0+{\textstyle{x^0\over 2}}\right) \,
 \tilde{\psi}_{a'}^* \left({\bf K}+{\bf k}',X^0-{\textstyle{x^0\over 2}}
 \right) = 
 \label{39a}\\
 &&\int d^3X\, d^3x\,
 \nonumber\\
 &&\quad 
 \psi_a \left( X+{\textstyle{x\over 2}}\right) \,
 e^{i \left[ \omega({\bf K}+{\bf k}) \left(X^0+\frac{1}{2}x^0 \right)
           - ({\bf K}+{\bf k})\cdot \left({\bf X}+\frac{1}{2}{\bf x}\right) 
      \right] }\,
 \nonumber\\
 &&\quad
 \psi_{a'}^* \left( X-{\textstyle{x\over 2}}\right)
 e^{-i \left[ \omega({\bf K}+{\bf k}') \left(X^0-\frac{1}{2}x^0 \right)
            - ({\bf K}+{\bf k}')\cdot \left({\bf X}-\frac{1}{2}{\bf x}
            \right) 
       \right] } \, ,
 \nonumber
 \end{eqnarray}
 \begin{eqnarray}
 &&
 \tilde{\psi}_b \left({\bf K}-{\bf k},Y^0+{\textstyle{y^0\over 2}}\right) \,
 \tilde{\psi}_{b'}^* \left({\bf K}-{\bf k}',Y^0-{\textstyle{y^0\over 2}}
 \right) = 
 \label{39b}\\
 &&\int d^3Y\, d^3y\,
 \nonumber\\
 &&\quad
 \psi_b \left(Y+{\textstyle{y\over 2}}\right) \,
 e^{i \left[ \omega({\bf K}-{\bf k}) \left(Y^0+\frac{1}{2}y^0 \right)
           - ({\bf K}-{\bf k})\cdot \left({\bf Y}+\frac{1}{2}{\bf y} \right) 
      \right] }\,
 \nonumber\\
 &&\quad
 \psi_{b'}^*\left(Y-{\textstyle{y\over 2}}\right)
 e^{-i \left[ \omega({\bf K}-{\bf k}') \left(Y^0-\frac{1}{2}y^0 \right)
            - ({\bf K}-{\bf k}')\cdot \left({\bf Y}-\frac{1}{2}{\bf y}
            \right) 
       \right] } \,.
 \nonumber
 \end{eqnarray}
 \end{mathletters}
The phase in (\ref{37}) reduces to
 \begin{eqnarray}
  && \Omega ({\bf K}, {\bf k})t_a - \Omega ({\bf K}, {\bf k'})t_{a'} =
 \label{40}\\
  && \ \ X^0\big[ \omega  ({\bf K}+{\bf k'})-\omega ({\bf K}+{\bf k})
                   +\omega  ({\bf K}-{\bf k'})-\omega ({\bf K}-{\bf k})\big]
 \nonumber \\
  && + x^0\Big[ 2K^0
 \nonumber\\
  && \ \ -{\textstyle{1\over 2}} 
  [\omega  ({\bf K}+{\bf k'})+\omega ({\bf K}+{\bf k})
  +\omega  ({\bf K}-{\bf k'})+\omega ({\bf K}-{\bf k})] \Big]
 \, ,
 \nonumber
 \end{eqnarray}
while the $\theta$-functions become
 \begin{eqnarray}
 && \theta \left(X^0-Y^0+{\textstyle{x^0-y^0 \over 2}} \right)
    \theta \left(X^0-Y^0-{\textstyle{x^0-y^0 \over 2}} \right) 
 \nonumber \\
 && = \int \frac{d\omega }{2\pi }\, 
 \frac{1}{\omega + i \epsilon }\,
  e^{-i \omega \left[ X^0-Y^0+\frac{1}{2} (x^0-y^0) \right] }\,
 \nonumber \\
 && \times\int \frac{d\omega '}{2\pi } \,
 \frac{1}{\omega '- i \epsilon } \,
  e^{i \omega '\left[ X^0-Y^0-\frac{1}{2} (x^0-y^0) \right] }
 \, .
 \label{41}
 \end{eqnarray}

We now define the single particle Wigner density $S(X,K)$ of the
source as 
 \begin{eqnarray}
  S(X,K) &=& \int d^4x\, e^{i K\cdot x}\,
  \rho\left(X^0+{\textstyle{x^0\over 2}},X^0-{\textstyle{x^0\over 2}}\right)
 \nonumber\\
  && \times\  
  \sum_{aa'} \nu_{aa'}\,
  \psi_a\left(X+{\textstyle{x\over 2}}\right) \,
  \psi_{a'}^*\left(X-{\textstyle{x\over 2}}\right) \, .
 \label{42}
 \end{eqnarray}
Using the hermiticity of the density matrix one easily shows that
$S(X,K)$ real. Then we can combine Eqs.~(\ref{39})-(\ref{42}) to
rewrite (\ref{37}) as 
 \begin{mathletters}
 \label{43}
 \begin{eqnarray}
  && P^{(aa)}({\bf p}_a,{\bf p}_b) = \int \frac{d^4 k}{(2\pi )^4} \,
       \frac{d^4 k'}{(2\pi )^4} \,
  \Phi\left({\textstyle{{\bf q}\over 2}},{\bf k}'\right) \,
  \Phi^*\left({\textstyle{{\bf q}\over 2}},{\bf k}\right) \,
 \nonumber\\
  &&\times \
  \frac{1}{K^0-k^0-\omega({\bf K}-{\bf k}) + i \epsilon } \,
  \frac{1}{K^0-{k^0}'-\omega({\bf K}-{\bf k}') - i \epsilon } \,
 \nonumber \\
   &&\times
   \int d^4X\, d^4Y \, e^{i (k-k')\cdot (X-Y)}\,
 \nonumber\\
   && \qquad \times\ 
   S\left(X,K+{\textstyle{k+k'\over 2}}\right)\,
   S\left(Y,K-{\textstyle{k+k'\over 2}}\right)\, .
 \label{43a}
 \end{eqnarray}
To obtain this expression we shifted the integration variables in
(\ref{41}) by defining $k^0=K^0-\omega -\omega ({\bf K}-{\bf k})$,
${k^0}' = K^0-\omega'-\omega ({\bf K}-{\bf k}')$.

The second diagonal term $P^{(bb)}$ can be calculated similarly:
 \begin{eqnarray}
  && P^{(bb)}({\bf p}_a,{\bf p}_b) = \int \frac{d^4 k}{(2\pi )^4} \,
       \frac{d^4 k'}{(2\pi )^4} \,
  \Phi\left({\textstyle{{\bf q}\over 2}},{\bf k}'\right) \,
  \Phi^*\left({\textstyle{{\bf q}\over 2}},{\bf k}\right) \,
 \nonumber \\
  && \times\ 
  \frac{1}{K^0+k^0-\omega ({\bf K}+{\bf k}) + i \epsilon } \,
  \frac{1}{K^0+{k^0}'-\omega ({\bf K}+{\bf k}') - i \epsilon } \,
 \nonumber \\
  &&\times
   \int d^4X\, d^4Y \, e^{i (k-k')\cdot (X-Y)}\,
 \nonumber \\
  && \qquad\times\ 
   S\left(X,K+{\textstyle{k+k'\over 2}}\right)\,
   S\left(Y,K-{\textstyle{k+k'\over 2}}\right)\, .
 \label{43b}
 \end{eqnarray}
By relabelling the integration variables,
 $$
 X\ \rightleftharpoons \ Y \ \ \ {\rm and} \ \ \
 k\ \rightarrow \ -k\ , \ \ k'\ \rightarrow \ -k' \, ,
 $$
and taking into account that $\Phi\left(\frac{1}{2} {\bf q},-{\bf
 k}\right) = \pm \Phi\left(\frac{1}{2} {\bf q},{\bf k}\right)$ for
bosons and fermions, respectively, one easily checks the equality
\begin{equation}
  P^{(aa)}({\bf p}_a,{\bf p}_b) = P^{(bb)}({\bf p}_a,{\bf p}_b) \, .
 \label{43c}
 \end{equation}

The cross terms $P^{(ab)}$ and $P^{(ba)}$ represent interference between
amplitudes of opposite time ordering. They va\-nish by causality.
Indeed, each cross term contains under the integral the product
of two retarded propagators, e.g. $1/[K^0-k^0-\omega ({\bf K}-{\bf k})
+ i \epsilon]$ and $1/[K^0+{k^0}'-\omega ({\bf K}+{\bf k}') - i
\epsilon]$ in $P^{(ab)}$, which in the complex $k^0$ and ${k^0}'$
planes have poles on the same side of the real $k^0$ and ${k^0}'$
axes while the corresponding time-energy exponents in the plane wave
factor have opposite signs. This latter fact means that when doing the
$k^0$ and ${k^0}'$ integrations, the corresponding contours must be
closed in opposite half planes, thus always missing one of the poles. 

Altogether we thus find 
\begin{equation}
 P_2({\bf p}_a,{\bf p}_b) = 2 P^{(aa)}({\bf p}_a,{\bf p}_b) \, .
\label{43d}
\end{equation}
\end{mathletters}
Equations (\ref{43}) are the main result of this paper. In the
following section we discuss it further and study 
various approximations. At this point let us only note that if we
neglect the phase factor resulting from the free propagation of the
first emitted particle until emission of the second one, i.e. use
Eq.~(\ref{32}) for the two-particle momentum amplitude, we obtain a
very similar expression to (\ref{43a}) where only the terms
$\omega ({\bf K}-{\bf k})$ resp. $-\omega ({\bf K}-{\bf k}')$
in the two energy denominators are missing. The consequences will be
discussed below. 

%%%%%%%%%%%%%%%%%%%%%%%%%%%%%%%%%%%%%%%%%%%%%%%%%%%%%%%%%%%%%%%%%%%%%
\section{Discussion and Approximations}
\label{sec3}
%%%%%%%%%%%%%%%%%%%%%%%%%%%%%%%%%%%%%%%%%%%%%%%%%%%%%%%%%%%%%%%%%%%%%

Expression (\ref{43a}) for the two-particle spectrum is not very
convenient in practice, because of the poles resulting from the two
energy denominators. Their contribution can, however, be evaluated
analytically by doing the $k^0$, ${k^0}'$ integrations using residue
calculus. For $X^0-Y^0>0$ one can close the two integration contours
in such a way that both poles are encircled and contribute. One then
obtains 
 \begin{mathletters}
 \label{44}
 \begin{eqnarray}
  && P_2({\bf p}_a,{\bf p}_b) =
  2 \int \frac{d^3 k}{(2\pi )^3} \, \frac{d^3 k'}{(2\pi )^3} \,
  \Phi\left({\textstyle{{\bf q}\over 2}},{\bf k}'\right) \,
  \Phi^*\left({\textstyle{{\bf q}\over 2}},{\bf k}\right) \,
 \nonumber \\
  && \qquad 
  \int d^4X\, d^4Y \, \theta (X^0-Y^0)\, e^{i (k-k')\cdot (X-Y)}\,  
 \nonumber \\
  && \qquad \times\ 
   S\left(X,K+{\textstyle{k+k'\over 2}}\right)\,
   S\left(Y,K-{\textstyle{k+k'\over 2}}\right)\, ,
 \label{44a}
 \end{eqnarray}
where
 \begin{equation}
   k^0  = K^0 - \omega({\bf K}-{\bf k}) \, , \quad
   {k^0}' = K^0 - \omega({\bf K}-{\bf k}') 
\label{44b}
\end{equation}
if the free time evolution between emission points is included, while
 \begin{equation}
   k^0 = {k^0}' = K^0 \ ,
 \label{44c}
 \end{equation}
 \end{mathletters}
if it is neglected.

%%%%%%%%%%%%%%%%%%%%%%%%%%%%%%%%%%%%%%%%%%%%%%%%%%%%%%%%%%%%%%%%%%%%%
\subsection{Wigner representation}
\label{sec3a}
%%%%%%%%%%%%%%%%%%%%%%%%%%%%%%%%%%%%%%%%%%%%%%%%%%%%%%%%%%%%%%%%%%%%%

Equation (\ref{43}) can be written as a folding relation between
Wigner densities. We define $p=\frac{1}{2} (k+k')$, $Q=k-k'$, and
 \begin{mathletters}
 \label{45}
 \begin{equation}
   \chi_{{\bf q}/2}(p) \equiv 
   \Phi\left({\textstyle{{\bf q}\over 2}},{\bf p}\right) \,
   \frac{1}{p^0-K^0+\omega ({\bf p}-{\bf K}) + i\epsilon}
 \label{45a}
 \end{equation}
for the case where the propagation between emission points is included, or
 \begin{equation}
   \chi_{{\bf q}/2}(p) \equiv
   \Phi\left({\textstyle{{\bf q}\over 2}},{\bf p}\right) \,
   \frac{1}{p^0-K^0 + i\epsilon}
 \label{45b}
 \end{equation}
 \end{mathletters}
if free propagation between emission points is neglected. The
two-particle cross section (\ref{43}) then reads 
 \begin{eqnarray}
  P_2({\bf p}_a,{\bf p}_b) &=& \int \frac{d^4 p}{(2\pi )^4} \, d^4X\, d^4Y\,
  S(X,K+p )
 \nonumber\\
  &&\quad \times\ W_{{\bf q}/2}(X-Y,p)\, S(Y,K-p) \, ,
 \label{46}
 \end{eqnarray}
with
 \begin{eqnarray}
   W_{{\bf q}/2}(X-Y,p) &=&  \int \frac{d^4 Q}{(2\pi )^4} \,
   e^{-i Q\cdot (X-Y)}\,
 \nonumber\\
   &\times& \chi_{{\bf q}/2}\left(p+{\textstyle{Q\over 2}}\right) \,
   \chi^*_{{\bf q}/2}\left(p-{\textstyle{Q\over 2}}\right) \, .
 \label{47}
 \end{eqnarray}
This function $W$ can be interpreted as the Wigner density associated
with the distorted wave describing the final state interactions. It
describes the power of the FSI to ``push'' two particles, which were
originally emitted with momenta $p'_a=K+p$ and $p'_b=K-p$, to the
observed values $p_a$ and $p_b$ (resp. $K$ and $q$). It is determined
by the ``modified distorted waves'' $\chi$. Please note that the
representation (\ref{46}) is generic; it was previously derived by
Pratt in Eq.~(2.8) of Ref.~\cite{pratt95} with quite different
methods. Different approximations in dealing with the propagation of
the first emitted particle between the emission points only result in
different ``modification factors'' associated with the distorted waves
$\Phi\left({{\bf q}\over 2},{\bf p}\right)$. Equations (\ref{45a}) and
(\ref{45b}) are two specific examples: we expect that different
assumptions about what happens to the first particle while it waits
for the appearance of the second one can be similarly included in
Eq.~(\ref{47}) by simply changing the ``modification factor'' in the
definition of $\chi_{{\bf q}/2}(p)$. 

%%%%%%%%%%%%%%%%%%%%%%%%%%%%%%%%%%%%%%%%%%%%%%%%%%%%%%%%%%%%%%%%%%%%%
\subsection{Free particle limit}
\label{sec3b}
%%%%%%%%%%%%%%%%%%%%%%%%%%%%%%%%%%%%%%%%%%%%%%%%%%%%%%%%%%%%%%%%%%%%%

To discuss the limiting case of no final state interactions it is best
to start from Eq.~(\ref{44a}). In this case the wave functions
$\phi_{\pm {\bf q}/2}({\bf r})$ in (\ref{16}) describing the relative
motion of the two particles on their way to the detector become plane
waves, with the momentum space representation
 \begin{equation}
   \Phi\left(\frac{\bf q}{2} ,{\bf k}\right) =
   {(2\pi)^3\over \sqrt{2}} \left[
   \delta^{(3)} \left({\textstyle{{\bf q}\over 2}}-{\bf k}\right) \pm
   \delta^{(3)} \left({\textstyle{{\bf q}\over 2}}+{\bf k}\right) 
   \right] \, .
 \label{48}
 \end{equation}
The two $\Phi$-functions in (\ref{44a}) result in four terms two of which
correspond to $\frac{1}{2} ({\bf k}+{\bf k}')= \pm {\bf q}$, ${\bf
  k}-{\bf k}'= 0$, while the other two terms have $\frac{1}{2} ({\bf
  k}+{\bf k}')= 0$, ${\bf k}-{\bf k}'= \pm {\bf q}$. The corresponding
energy values depend on whether or not we include free propagation
between the emission points. If we include it, the energies are
$\frac{1}{2} (k^0+{k^0}')=\pm q^0$, $k^0-{k^0}'=0$ for the first two terms
and $\frac{1}{2} (k^0+{k^0}')=0$, $k^0-{k^0}'=\pm q^0$ for the last two
terms ($q^0=E_a-E_b$). The two-particle cross section becomes
 \begin{eqnarray}
   P_2({\bf p}_a,{\bf p}_b) &=& \int d^4X\, d^4Y\, \theta (X^0-Y^0)\,
 \nonumber\\
   &\times& \bigg\{ 
   S\left(X,K{+}{\textstyle{q\over2}}\right)\,
   S\left(Y,K{-}{\textstyle{q\over2}}\right) 
 \nonumber\\
   && +
   S\left(X,K{-}{\textstyle{q\over2}}\right)\,
   S\left(Y,K{+}{\textstyle{q\over2}}\right) 
 \nonumber \\
   \pm \Bigl[ e^{i q\cdot (X-Y)} &{+}& e^{- i q\cdot (X-Y)} \Bigr]\,
   S(X,K) \, S(Y,K) \bigg\} \, .
 \label{49}
 \end{eqnarray}
After relabelling $X\leftrightarrow Y$ in the second and fourth term
and using $\theta(X^0{-}Y^0)+\theta(Y^0{-}X^0)=1$ we obtain
 \begin{eqnarray}
   P_2({\bf p}_a,{\bf p}_b) &=& \int d^4X\, d^4Y\, \Bigl[
   S\left(X,K{+}{\textstyle{q\over2}}\right)\,
   S\left(Y,K{-}{\textstyle{q\over2}}\right) 
 \nonumber\\
   &&\quad \pm e^{i q\cdot (X-Y)} \, S(X,K)\, S(Y,K ) \Bigr]\, .
 \label{50}
 \end{eqnarray}
After normalization with the single particle spectra we thus recover
the correct expression (\ref{1}) for the correlator in the limit of
vanishing final state interactions.

If the free propagation between emission points is neglected, according
to (\ref{44c}) all four terms have the same energies 
$\frac{1}{2} (k^0+{k^0}')=K^0$, $k^0-{k^0}'=0$, and it is obvious that
it is not possible to recover an expression with any close similarity
to (\ref{50}). The phases resulting from the time evolution of the
first particle until the appearance of the second one are thus crucial
for the correct free-particle limit. One may be able to modify the
propagation between emission points, but not to drop it completely.
The correct free particle limit requires the appearance of a phase
$\sim (t_a-t_b)$.

%\input fsi2.tex

%%%%%%%%%%%%%%%%%%%%%%%%%%%%%%%%%%%%%%%%%%%%%%%%%%%%%%%%%%%%%%%%%%%%%
\subsection{The smoothness approximation}
\label{sec3c}
%%%%%%%%%%%%%%%%%%%%%%%%%%%%%%%%%%%%%%%%%%%%%%%%%%%%%%%%%%%%%%%%%%%%%

For practical applications one would like to know how to combine this
formalism with classical event generators which determine the emission
functions $S(x,K)$ as a set of last interaction points in the kinetic
evolution the many-particle collision system. For particles without
FSI it was shown in \cite{zhang97} that the symmetrized two particle
cross section (\ref{50}) can be computed by sampling the generated
particles at the on-shell momenta $({\bf p}_a,E_a)$, $({\bf p}_b,E_b)$
for the direct term, and at the on-shell value $({\bf K},E_K)$ for the
exchange term (with both selected particles having the {\em same} 
on-shell momentum $({\bf K},E_K)$!). The pairs selected for the
exchange term are then multiplied with a ``Bose-Einstein weight factor''
$\cos {q\cdot (x_i-x_j)}$ where $q$ is the relative momentum value at
which we wish to know the correlator ({\it not} that of the selected
pairs which have $p_i=p_j=K$), and $x_i-x_j$ is the space-time distance
between the particles in the selected pairs. This does no longer work
in the presence of FSI. Equations (\ref{44a}), (\ref{46}) show that
now we need to sample the emission function $S(x,p)$ at all possible
and in general off-shell values of the momentum $p$. The particles
become only on-shell when they reach the detector, by virtue of the
FSI. It is thus basically impossible to simulate Eqs.~(\ref{44a}),
(\ref{46}) with classical event generators unless one imposes a
further approximation which essentially eliminates all off-shell
effects. 

This step is known as the smoothness approximation in which
one assumes that $S(x,K\pm p)$ has a sufficiently weak momentum
dependence that, over the $p$-range where $W_{{\bf q}/2}(X-Y,p)$
in Eq.~(\ref{46}) is non-vanishing, $S(x,K\pm p)$ can be replaced by
either $S(x,K)$ or $S(x,K\pm p)$, depending on what seems more
convenient. The domain of support of the function $W_{{\bf q}/2}(X-Y,p)$ 
is a measure of the ability of the FSI to change the particle momenta
after freeze-out. For our purposes the smoothness approximation can be
considered reliable as long as the typical shift in momentum in the
FSI is small compared to the ${\bf q}$-range over which the correlator
shows interesting structure. Note that for massive particles the FSI
cause mostly a change of the spatial momentum while the corresponding
energy transfer is very small. This is the basic reason why off-shell
effects are small and why the smoothness approximation works.

Our implementation of the smoothness approximation differs
in a crucial detail from previous approaches
\cite{pratt86,pratt90,pratt95,lednicky}. By formulating the
approximation concisely and implementing it systematically we obtain
an expression which shows a strong similarity to the free particle
case in that it evaluates for the direct and exchange terms the source
function at different momenta. In this way we automatically avoid 
some of the possible pathologies in the behaviour of the correlator
\cite{zhang97,martin97,pratt97} which may arise from the more
traditional versions of the smoothness approximation.

We start from Eq.~(\ref{44}) which we rewrite as
 \begin{eqnarray}
  && P_2({\bf p}_a,{\bf p}_b) =
  2 \int d^4x\, d^4y\, \theta(y^0)
 \nonumber\\
  &&\quad\times
  \int \frac{d^3 k}{(2\pi )^3} \, \frac{d^3 k'}{(2\pi )^3} \,
  \Phi\left({\textstyle{{\bf q}\over 2}},{\bf k}'\right) \,
  \Phi^*\left({\textstyle{{\bf q}\over 2}},{\bf k}\right) \,
 \nonumber\\
  &&\quad\times\ 
  e^{-i[\omega({\bf K-k})-\omega({\bf K-k}')] y^0}\,  
  e^{-i ({\bf k}-{\bf k}')\cdot {\bf y}}\,  
 \nonumber \\
   &&\quad \times\ 
   S\left(x+{\textstyle{y\over 2}},K+{\textstyle{k+k'\over 2}}\right)\,
   S\left(x-{\textstyle{y\over 2}},K-{\textstyle{k+k'\over 2}}\right)\, .
 \label{50+1}
 \end{eqnarray}
Let us expand the frequency difference in the temporal phase factor 
for small values of ${\bf k,k}'$, keeping terms up to second order:
 \begin{eqnarray}
  && \left[\omega({\bf K}-{\bf k})-\omega ({\bf K}-{\bf k}')\right] 
 \nonumber \\
  && \approx - {1\over E_K} ({\bf k} - {\bf k}')
   \cdot \left({\bf K} - {{\bf k}+{\bf k}'\over 2}\right)\, .
 \label{50+2}
 \end{eqnarray}
Using (\ref{36}), the two-particle spectrum thus becomes
 \begin{eqnarray}
  && P_2({\bf p}_a,{\bf p}_b) =
  \int d^4x\, d^4y\, \theta(y^0)
 \nonumber\\
  &&\ \times \int \frac{d^3 k}{(2\pi )^3} \, \frac{d^3 k'}{(2\pi )^3} \,
  e^{-i ({\bf k}-{\bf k}')\cdot \left({\bf y} -{1\over E_K}
      \left({\bf K} - {{\bf k}+{\bf k}'\over 2} \right) y^0\right)}\,  
\nonumber\\
  &&\ \times \Bigl[
  \tilde\phi^*\left({\textstyle{{\bf q}\over 2}},{\bf k}\right) \,
  \tilde\phi\left({\textstyle{{\bf q}\over 2}},{\bf k}'\right) \, \pm
  \tilde\phi^*\left({\textstyle{{\bf q}\over 2}},{\bf k}\right) \,
  \tilde\phi\left(-{\textstyle{{\bf q}\over 2}},{\bf k}'\right) \,
 \nonumber\\
  &&\ \ \pm
  \tilde\phi^*\left(-{\textstyle{{\bf q}\over 2}},{\bf k}\right) \,
  \tilde\phi\left({\textstyle{{\bf q}\over 2}},{\bf k}'\right) \, +
  \tilde\phi^*\left(-{\textstyle{{\bf q}\over 2}},{\bf k}\right) \,
  \tilde\phi\left(-{\textstyle{{\bf q}\over 2}},{\bf k}'\right) \Bigr]
 \nonumber \\
   &&\ \times\ 
   S\left(x+{\textstyle{y\over 2}},K+{\textstyle{k+k'\over 2}}\right)\,
   S\left(x-{\textstyle{y\over 2}},K-{\textstyle{k+k'\over 2}}\right)\, .
 \label{50+3}
 \end{eqnarray}
The functions $\tilde\phi$ describe the probability amplitude of
finding in the FSI distorted wave with asymptotic relative momentum 
$\pm {\bf q}/2$ a plane wave with momentum ${\bf k}$ resp. ${\bf
  k}'$. These functions are peaked around ${\bf k},{\bf k}' = \pm {\bf
  q}/2$, and the peaking is the stronger the weaker the final 
state interactions are. The four terms in the square bracket in
(\ref{50+3}) thus peak at $(k+k')/2= q/2$, 0, 0, and $-q/2$,
respectively. (One easily checks that these equations hold for the
corresponding 4-vectors.) We will use this property by replacing
$(k+k')/2$ in the phase factor and in the arguments of the emission 
functions by the corresponding peak locations. This allows to pull the
emission functions outside the ${\bf k,k}'$ integrations; it also
removes the quadratic dependence of the phase factor on ${\bf k,k}'$
in favor of a linear one, turning the integrations over ${\bf k}$ and
${\bf k}'$ into normal Fourier integrals. The Fourier integrals can be
performed, giving the corresponding relative wave functions in
coordinate space: 
 \begin{eqnarray}
  && P_2({\bf p}_a,{\bf p}_b) =
  \int d^4y\, \theta(y^0) 
 \label{50+4}\\
  &&\left[ 
  \left\vert \phi_{{\bf q}/2}({\bf y}{-}{\bf v}_b y^0) \right\vert^2
  \int d^4x\,S\left(x+{\textstyle{y\over 2}},p_a\right)\,
             S\left(x-{\textstyle{y\over 2}},p_b\right) \right.
 \nonumber\\
  &&\  \pm 2\, {\rm Re}
  \left( \phi^*_{{\bf q}/2}({\bf y}{-}{\bf v} y^0) 
         \phi_{-{\bf q}/2}({\bf y}{-}{\bf v} y^0) \right)\,
 \nonumber\\
  && \qquad \qquad \qquad \times
  \int d^4x\,S\left(x+{\textstyle{y\over 2}},K\right)\,
             S\left(x-{\textstyle{y\over 2}},K\right) 
 \nonumber\\
  && + \left.
  \left\vert \phi_{-{\bf q}/2}({\bf y}{-}{\bf v}_a y^0) \right\vert^2
  \int d^4x\,S\left(x+{\textstyle{y\over 2}},p_b\right)\,
             S\left(x-{\textstyle{y\over 2}},p_a\right) \right] . 
 \nonumber
 \end{eqnarray}
Here we defined the three velocities
 \begin{equation}
   {\bf v} = {{\bf K}\over E_K}\, ,\quad
   {\bf v}_a = {{\bf p}_a\over E_K}\, ,\quad
   {\bf v}_b = {{\bf p}_b\over E_K}
 \label{50+5}
 \end{equation}
associated with the observed particle momenta ${\bf p}_a$, ${\bf p}_b$, 
and their average ${\bf K}$. 

Using the relations (\ref{18}) one shows that in the middle term in
(\ref{50+4}) the factor containing the FSI relative wave functions is
even under a simultaneous sign change of ${\bf y}$ and $y^0$. The
product of emission functions in this term is even under the same sign
change, too, which means that in this term we can replace $\theta
(y^0)$ by $\frac{1}{2} [\theta (y^0)+\theta (-y^0)]=\frac{1}{2}$. The
first and last term in (\ref{50+4}) do not have such a symmetry, but
can be combined by substituting $y\to -y$ in the last term to yield
the final expression 
 \begin{eqnarray}
  && P_2({\bf p}_a,{\bf p}_b) =
  \int d^4x\,d^4y\,S\left(x+{\textstyle{y\over 2}},p_a\right)\,
       S\left(x-{\textstyle{y\over 2}},p_b\right) 
 \nonumber\\
  &&\ \times \left[ \theta(y^0) 
  \left\vert \phi_{{\bf q}/2}({\bf y}{-}{\bf v}_b y^0) \right\vert^2
  + \theta(-y^0) 
  \left\vert \phi_{{\bf q}/2}({\bf y}{-}{\bf v}_a y^0) \right\vert^2
  \right]
 \nonumber\\
  &&\pm \int d^4x\,d^4y\,S\left(x+{\textstyle{y\over 2}},K\right)\,
       S\left(x-{\textstyle{y\over 2}},K\right)
 \nonumber\\
  && \qquad \quad \times\ 
     \phi^*_{-{\bf q}/2}({\bf y}{-}{\bf v} y^0) \, 
     \phi_{{\bf q}/2}({\bf y}{-}{\bf v} y^0) .
 \label{50+6}
 \end{eqnarray}
Once again one checks that in the limit of vanishing FSI (i.e. by
replacing the functions $\phi$ by plane waves), and using the
approximation $K^0\approx E_K$ in accordance with the approximation
made in Eq.~(\ref{50+2}), the correct expression (\ref{50}) is
recovered. 

It is worthwhile to discuss the physical meaning of the arguments at
which the FSI distorted waves $\phi$ in (\ref{50+6}) must be
evaluated. The sketch presented in Fig.~\ref{F1} shows the relevant 
possibilities for the first term in (\ref{50+6}) (the ``direct
term''). In this term the two emission functions are evaluated at the
observed momenta $p_a,p_b$, corresponding to particle velocities ${\bf
  v}_a, {\bf v}_b$. Fig.~\ref{F1} shows that in each case (i.e. for
both possible time orderings between the two emission points) the
argument of the FSI distorted wave $\phi$ corresponds to the spatial
distance of the two particles {\em at the time when the second
  particle freezes out}, i.e. when the {\em pair} first exists. The
different velocities which arise for the two different time orderings
reflect the velocity of the earlier emitted particle in each case. In
the second term of (\ref{50+6}) (the ``exchange term'') the two
emission functions are evaluated at the {\em same} momentum, namely
the average pair momentum $K$, and correspondingly for both 
time orderings the earlier emitted particle has the velocity
${\bf v}$ corresponding to this momentum. Again the argument of the
FSI distorted wave is the spatial distance between the two particles
at the time of emission of the second particle.

%%%%%%%%%%%%%%%%%%%%%%%%%%%%%%%%%%%%%%%%%%%%%%%%%%%%%%%%%%%%%%%%%%%%%
\subsection{Implementation in event generators}
\label{sec3d}
%%%%%%%%%%%%%%%%%%%%%%%%%%%%%%%%%%%%%%%%%%%%%%%%%%%%%%%%%%%%%%%%%%%%%

Equation~(\ref{50+6}) can be easily implemented into event generators,
following essentially the same procedure as given in
Ref.~\cite{zhang97}:

For the {\em direct term} one selects all pairs $(i,j)$ with $p_i=p_a$, 
$p_j=p_b$ within a given numerical accuracy (bin width) which is 
essentially dictated by event statistics. Each pair is multiplied with 
a weight given by the corresponding probability density $\vert 
\phi_{{\bf q}/2}\vert^2$ of the FSI distorted wave. The latter must be 
evaluated in a frame in which the pair moves non-relativistically, 
best in the pair rest frame where ${\bf K} = ({\bf p}_a + {\bf p}_b)/2 
=0$. (Then $E_K=m$, and the velocities (\ref{50+5}) reduce to their 
usual nonrelativistic definition.) From the space-time coordinates 
$x_i,x_j$ of the pair in the event generator frame one calculates the 
distance ${\bf y}^*_{ij}$ between the two particles in the pair rest
frame at the time when the second particle is produced. One then
computes $\vert \phi_{{\bf q}^*/2}({\bf y}^*_{ij})\vert^2$ and weights
the selected pair $(i,j)$  
with this number. In this expression ${\bf q}^*$ is the spatial 
relative momentum between the two particles in the pair rest frame 
which must be computed from $p_a,p_b$ in the event generator frame. 
(We remark that in the absence of FSI, the corresponding weight is 
simply 1.) The complete direct term is obtained by summing over all 
such pairs.  

For the {\em exchange term}, the selection of pairs and weights is a
little less intuitive \cite{zhang97}: One selects all pairs $(i,j)$ with 
$p_i=p_j=K$ (i.e. {\em on-shell particles} (!) with ${\bf p}_i={\bf 
 p}_j = {\bf K}$ and $E_i=E_j=E_K$), again within the same numerical 
accuracy (bin width) as above. From the production coordinates 
$x_i,x_j$ one again computes the spatial distance ${\bf y}^*_{ij}$ 
between the two particles in the pair at the time of emission of the 
second one, in the pair rest frame ${\bf K}=0$. This distance is used 
to compute the weight $\phi^*_{-{\bf q}^*/2}({\bf  y}^*_{ij})
 \phi^*_{{\bf q}^*/2}({\bf  y}^*_{ij})$ for this pair. The value of
${\bf q}^*$ here is {\em the same as above in the direct term},
i.e. it is computed from $p_a$ and $p_b$ by transforming into the pair
rest frame, not from $p_i=p_j=K$. (Without FSI, the corresponding
weight would be $\cos({\bf q}^*\cdot{\bf y}_{ij}^*)$ \cite{zhang97}.)
The full exchange term is obtained by summing over all such pairs.  

Finally one must normalize the correlator by the product of single
particle spectra,
 \begin{equation}
 \label{single}
   P_1({\bf p}_a)\,P_1({\bf p}_b) = 
   \int d^4x\, S(x,p_a)\,\int d^4y\, S(y,p_b)\, .
 \end{equation}
This normalization is best obtained from the pairs selected for the
direct term above by multiplying them with unit weights. 

One may object to the use of event generators for the emission 
function because they fix particle momenta and coordinates 
simultaneously and thus violate the uncertainty principle. It was 
shown in \cite{zhang97} how to generate from an event generator a 
quantum mechanically consistent Wigner density $S(x,p)$ by folding the 
event generator output with minimum uncertainty wave packets. The 
corresponding quantum mechanically consistent algorithm for computing 
single- and two-particle spectra given in \cite{zhang97} is easily 
generalized to include FSI effects, by simply replacing the factors 
1 and $\cos({\bf q}^*\cdot{\bf y}^*_{ij})$ in the direct and exchange 
terms, respectively, by the correct FSI weights as discussed above.  

%%%%%%%%%%%%%%%%%%%%%%%%%%%%%%%%%%%%%%%%%%%%%%%%%%%%%%%%%%%%%%%%%%%%%%%%%%%%
\subsection{Comparison to previously published expressions}
\label{sec3e}
%%%%%%%%%%%%%%%%%%%%%%%%%%%%%%%%%%%%%%%%%%%%%%%%%%%%%%%%%%%%%%%%%%%%%%%%%%%%

In previous treatments \cite{pratt86,pratt90,pratt95,pratt97} the 
smoothness approximation was implemented in a different way: the 
smoothness of $S(x,p)$ as a function of momentum was used to replace
the momentum arguments of the two emission functions in both the 
direct and the exchange term of Eqs.~(\ref{50}), (\ref{50+6}) in exactly 
the same way, namely either by $p_a$ in the first and by $p_b$ in the 
second, or by $K$ in both emission functions. The first of these two 
alternatives can, for sources with very strong $x$-$p$-correlations, 
lead to correlators with pathological behaviour, as discussed in 
\cite{zhang97,martin97,pratt97}. The second alternative was 
exploited in the FSI studies presented in Refs.
\cite{pratt86,pratt90,pratt95}. In this case both the direct and the 
exchange term in (\ref{50+6}) involve the same combination of emission 
functions, namely 
 \begin{equation}
   D(y,K) \equiv \int d^4x \, 
   S\left( x+{\textstyle{y\over 2}},K\right) \,
   S\left( x-{\textstyle{y\over 2}},K\right) \, .
 \label{54}
 \end{equation}
$D(y,K)$ is the ``relative distance distribution'' of the source,
i.e. the distribution of relative space-time distances $y$ between the 
particles in emitted pairs with momentum $K$. Obviously $D$ is an even 
function of $y$, $D(y,K) = D(-y,K)$. Also, since in the direct term 
of (\ref{50+6}) the emission function arguments $p_a,p_b$ are now both 
replaced by $K$, the velocities ${\bf v}_a,{\bf v}_b$ in the 
corresponding FSI distorted wave factors must also both be replaced by 
${\bf v}$. The two time orderings can then be combined by using 
$\theta(y^0)+\theta(-y^0)=1$, and we obtain
 \begin{equation}
   P_2({\bf p}_a,{\bf p}_b) = P_2({\bf q},{\bf K}) \approx
   \int d^4y\, w(y;q,K)\, D(y,K) \, ,
 \label{55}
 \end{equation}
with the ``weight function'' 
 \begin{eqnarray}
   w(y;q,K) &=& 
   \left\vert \Phi_{{\bf q}/2}({\bf y}{-}{\bf v} y^0)\right\vert^2
 \label{56}\\
  &=& {1\over 2} \left\vert \phi_{{\bf q}/2}({\bf y}{-}{\bf v} y^0) \pm
     \phi_{{\bf q}/2}(-({\bf y}{-}{\bf v} y^0)) \right\vert^2 \, .
 \nonumber
 \end{eqnarray}
[Note that if $P_2$ is approximated as in (\ref{55}) one should also for
consistency use the corresponding approximation $P_1({\bf p}_a)
P_1({\bf p}_b) = \left[P_1({\bf K})\right]^2$ for the normalization of
the correlator.] Eqs.~(\ref{55}), (\ref{56}) are identical with 
Eq.~(2.11) in Ref.~\cite{pratt95}. In the pair rest system (${\bf 
v}=0={\bf K}$) $P_2$ reduces to 
 \begin{equation}
    P_2({\bf p}_a,{\bf p}_b) \approx \int d^3y\,
    \left\vert \Phi_{{\bf q}/2}({\bf y})\right\vert^2
    \int dy^0\, D(y,K) \, .
 \label{57}
 \end{equation}
This expression was first written down by Koonin \cite{koonin77} and 
has recently been used in \cite{brown97}. Note that on the r.h.s. of 
(\ref{57}) everything must be evaluated in the pair rest system with 
${\bf K}=0$. In this version of the smoothness approximation the 
two-particle spectrum $P_2({\bf q,K})$ is thus given by the time 
integrated relative distance distribution in the pair rest system 
defined by ${\bf K}$, weighted with the probability density of the 
outgoing distorted wave with relative momentum ${\bf q}$ at the 
spatial relative distance at which the pair was created.  

%%%%%%%%%%%%%%%%%%%%%%%%%%%%%%%%%%%%%%%%%%%%%%%%%%%%%%%%%%%%%%%%%%%%%%
\subsection{Nonidentical particles}
\label{sec3f}
%%%%%%%%%%%%%%%%%%%%%%%%%%%%%%%%%%%%%%%%%%%%%%%%%%%%%%%%%%%%%%%%%%%%%%

For correlations introduced by FSI between non-iden\-tical pairs one has
to take into account two differences: 

(1) the two masses $m_a$ and $m_b$ are unequal, and

(2) the two-particle wave-functions are not
sym\-me\-trized resp. antisymmetrized. \newline
The first modification requires the introduction of correspondingly
modified center-of-mass and relative coordinates in several steps of
the derivation given in Sec.~\ref{sec2}. Still, the problem remains
non-relativistic in any system in which the pair is at rest or moves
non-relativistically, and the rest of the derivation goes through as
before. The final result is again given by Eq.~(\ref{50+6}); the only
difference is that now $\Phi_{{\bf q}/2}({\bf y})$ is just the
distorted wave for the relative motion including FSI, not its
symmetrized form (\ref{36}). As a result of the mass asymmetry,
the weight function $w(y;q,K)$ even in the pair rest system will no 
longer be symmetric in ${\bf y}$. Its asymmetry will be reflected in 
an asymmetry of the correlator under sign change of ${\bf q}$ which 
can be used to extract information about the average distance between 
the effective sources from which the two particle species originate 
\cite{led96}.  

%%%%%%%%%%%%%%%%%%%%%%%%%%%%%%%%%%%%%%%%%%%%%%%%%%%%%%%%%%%%%%%%%%%%%%
\subsection{Coulomb final state interactions}
\label{sec3g}
%%%%%%%%%%%%%%%%%%%%%%%%%%%%%%%%%%%%%%%%%%%%%%%%%%%%%%%%%%%%%%%%%%%%%%

For completeness and to correct a few confusing typographical errors
in Ref.~\cite{pratt86} we discuss explicitly the distorted wave for
Coulomb final state interacions. In this case the solution of the 
Schr\"odinger Eq.~(\ref{16}) is 
 \begin{equation}
  \phi_{\pm{\bf q}/2} ({\bf r}) =
  \Gamma (1+ i \eta ) e^{-\frac{1}{2} \pi \eta }\,
  e^{\pm\frac{i}{2} {\bf q}\cdot {\bf r} }
  F(-i\eta;1;iz_\mp) \, ,
 \label{59}
 \end{equation}
where ($q=\vert {\bf q} \vert,\, r=\vert {\bf r}\vert$)
 \begin{equation}
 \label{60}
   \eta = {\alpha \mu \over q/2} = {\alpha m\over q}
 \end{equation}
is the Sommerfeld parameter, and
 \begin{equation}
 \label{61}
   z_\mp = {\textstyle{1\over 2}}\left( qr \mp {\bf q}\cdot{\bf r}\right)
   = {\textstyle{1\over 2}} qr (1 \mp \cos\theta) \, ,
 \end{equation}
$\theta$ being the angle between {\bf q} and {\bf  r}. $F(-i\eta;1;iz)$
is a confluent hypergeometric function with the series expansion
 \begin{eqnarray}
 \label{62}
   && F(-i\eta;1;iz) = \sum_{n=0}^\infty 
                      {\Gamma(n-i\eta)\over \Gamma(-i\eta)}\, 
                      {(iz)^n \over (n!)^2}
 \nonumber\\
   &&= 1 -i\eta \sum_{n=1}^\infty (1-i\eta)(2-i\eta)\cdots(n-1-i\eta)
                      {(iz)^n \over (n!)^2}\, .
 \end{eqnarray}
For small $\eta\ll 1$ this can be approximated by \cite{pratt86}
 \begin{eqnarray}
 \label{63}
   F(-i\eta;1;iz) &\approx& 1 -i\eta \sum_{n=1}^\infty 
                            {(iz)^n \over n\cdot n!}
 \nonumber\\
   &=& 1 - i\eta \Bigl( {\rm Ci}(z) + i\, {\rm Si}(z) - \ln z - \gamma \Bigr)
   \, .
 \end{eqnarray}
This is good for pions with $q \gg 1$ MeV and for protons with $q \gg
7$ MeV. The square of the normalization factor in (\ref{59}) yields the
well-known Gamov factor
 \begin{equation}
   G(q) \equiv {2\pi\eta \over e^{2\pi\eta} -1} =
   e^{-\pi\eta}\, \vert\Gamma(1+i\eta)\vert^2 \, .
 \label{64}
 \end{equation}
With these expressions the two-particle cross section (\ref{50+6})
takes the form
 \begin{mathletters}
 \label{65}
 \begin{equation}
   P_2({\bf q,K}) = G(q)\, 
   \Bigl[ I_1({\bf q,K}) \pm I_2({\bf q,K}) \Bigr] \, ,
 \label{65a}
 \end{equation}
where the direct term is given by
 \begin{eqnarray}
 \label{65b}
   I_1({\bf q,K}) &=& \int d^4y \, \Bigl[ 
   \theta(y^0)\,\left\vert F\left(-i\eta;1;iz_-^b(y^0)\right) \right\vert^2 
 \nonumber\\
   && \qquad + \theta(-y^0)\,
                \left\vert F\left(-i\eta;1;iz_-^a(y^0)\right) \right\vert^2 
   \Bigr]
 \\
    \times&{\int}& d^4x\, 
    S\left(x+{\textstyle{y\over 2}},K+{\textstyle{q\over 2}}\right)\,
    S\left(x-{\textstyle{y\over 2}},K-{\textstyle{q\over 2}}\right)
 \nonumber
 \end{eqnarray}
while the exchange term is
 \begin{eqnarray}
 \label{65c}
   I_2({\bf q,K}) 
   &=& \int d^4y \, e^{i{\bf q}\cdot({\bf y}-{\bf v}y^0)} \, D(y,K)
 \\
   &\times& 
   F\left(i\eta;1;-iz_+(y^0)\right)\,
   F\left(-i\eta;1;iz_-(y^0)\right)\, .
 \nonumber
 \end{eqnarray}
 \end{mathletters}
Here $D(y,K)$ is the relative distance distribution (\ref{54}), and 
$z_-^b(y^0)$, $z_-^a(y^0)$, $z_\pm(y^0)$ are determined with
the help of (\ref{61}) by substituting ${\bf r} \to {\bf y}{-}{\bf
  v}_b y^0, \, {\bf y}{-}{\bf v}_a y^0$, and ${\bf y}{-}{\bf v} y^0$,
respectively. Note that the integrals $I_{1,2}$ should be evaluated in
(or close to) the frame where ${\bf K}={\bf v}=0$.

Equations (\ref{65}) correct Eq.~(3.11) in Ref.~\cite{pratt86}.

%%%%%%%%%%%%%%%%%%%%%%%%%%%%%%%%%%%%%%%%%%%%%%%%%%%%%%%%%%%%%%%%%%%%%%%%%%%
\section{Conclusions}
\label{sec4}
%%%%%%%%%%%%%%%%%%%%%%%%%%%%%%%%%%%%%%%%%%%%%%%%%%%%%%%%%%%%%%%%%%%%%%%%%%%

Let us shortly summarize our main results:

We studied the effect of two-body final state interactions on the
two-particle coincidence cross section, both for pairs of identical
and of non-identical particles. We used only two approximations: (1)
the source is chaotic, and (2) the total particle multiplicity is
large. The first of these two assumptions is crucial since it allows
to factorize the two-particle Wigner density of the source and express
the two-particle cross section in terms of the single-particle Wigner
density, i.e. the emission function $S(x,p)$ of the source.

From these two assumptions we derived, without further approximation,
the general expression (\ref{44}) for the two-particle cross section
in terms of the emission function $S(x,p)$ and the FSI distorted
relative wave functions in momentum space, $\Phi\left({{\bf q}\over
  2},{\bf k}\right)$. This expression can be rewritten in the generic
Wigner representation (\ref{46}). In spite of its generality, this
expression is of limited practical usefulness because it involves the
emission function at arbitrary of-shell momenta which is usually not
known. Since for massive particles the FSI shifts, however, mostly the
spatial momenta of the particles while the corresponding energy
transfer is very small, it is in practice possible to replace the
off-shell momenta in the emission function by on-shell values. We do
this systematically in the framework of the so-called smoothness
approximation which we discussed in Sec.~\ref{sec3c}. Here our
treatment differs from previously published ones, and our formula
(\ref{50+6}) thus improves upon known results. 

Eq.~(\ref{50+6}) involves only on-shell momenta and can thus be
implemented in classical event generators. This was discussed in
Sec.~\ref{sec3d}. In contrast to previously published expressions, our
result exhibits the same asymmetry in the momentum arguments of the
emission functions between the direct and exchange terms as expression
(\ref{50}) for the free case to which it correctly reduces when the
FSI are switched off.

This asymmetry between the direct and exchange terms is eliminated
if one implements the more stringent version of the smoothness
approximation used by Pratt and collaborators which replaces the
momentum argument of the emission function by the average pair
momentum $K$ everywhere. In Sec.~\ref{sec3e} we showed how in this way
the well-known Koonin-Pratt formula, Eqs.~(\ref{55})-(\ref{57}), is
recovered.

The differences between the ``correct'' asymmetric form and the
``approximate'' symmetric Koonin-Pratt form were recently extensively
investigated for the case of free particle propagation without FSI
after freeze-out. They were found to be potentially severe for sources
with strong $x$-$p$-correlations (e.g. sources which feature strong
collective expansion) \cite{zhang97,martin97}, but the problem appears
to be less serious if the source is sufficiently large
\cite{pratt97}. While the existence of the ``correct'' asymmetric
expression (\ref{50+6}) (for Coulomb FSI an explicit expression was
given in (\ref{65})) now eliminates the need for using the approximate
and somewhat uncertain Koonin-Pratt formula, it would still be nice to
have a quantitative feeling for the error margin associated with the
Koonin-Pratt approximation. A systematic numerical study of this
question which includes FSI effects is under way. 

%%%%%%%%%%%%%%%%%%%%%%%%%%%%%%%%%%%%%%%%%%%%%%%%%%%%%%%%%%%%%%%%%%%%%%%%%%%

\acknowledgments

D.A. would like to thank P. Braun-Munzinger, D. Miskowiec, G. Zinovjev
and S. Voloshin for stimulating discussions. U.H. would like to thank
S. Pratt for his hospitality and many interesting discussions and 
useful suggestions regarding the manuscript. This work was supported
by the Deutsche Forschungsgemeinschaft (DFG), Bundesministerium f\"ur
Bildung und Forschung (BMBF), Gesellschaft f\"ur Schwerionenforschung
(GSI), and by grant INTAS-94-3941.  

%%%%%%%%%%%%%%%%%%%%%%%%%%%%%%%%%%%%%%%%%%%%%%%%%%%%%%%%%%%%%%%%%%%%%%%%%%%

%%%%%%%%%%%%%%%%%%%%%%%%%%%%%%%%%%%%%%%%%%%%%%%%%%%%%%%%%%%%%%%%%%%%
% Figures
%%%%%%%%%%%%%%%%%%%%%%%%%%%%%%%%%%%%%%%%%%%%%%%%%%%%%%%%%%%%%%%%%%%%

\vspace*{14cm}
\includegraphics{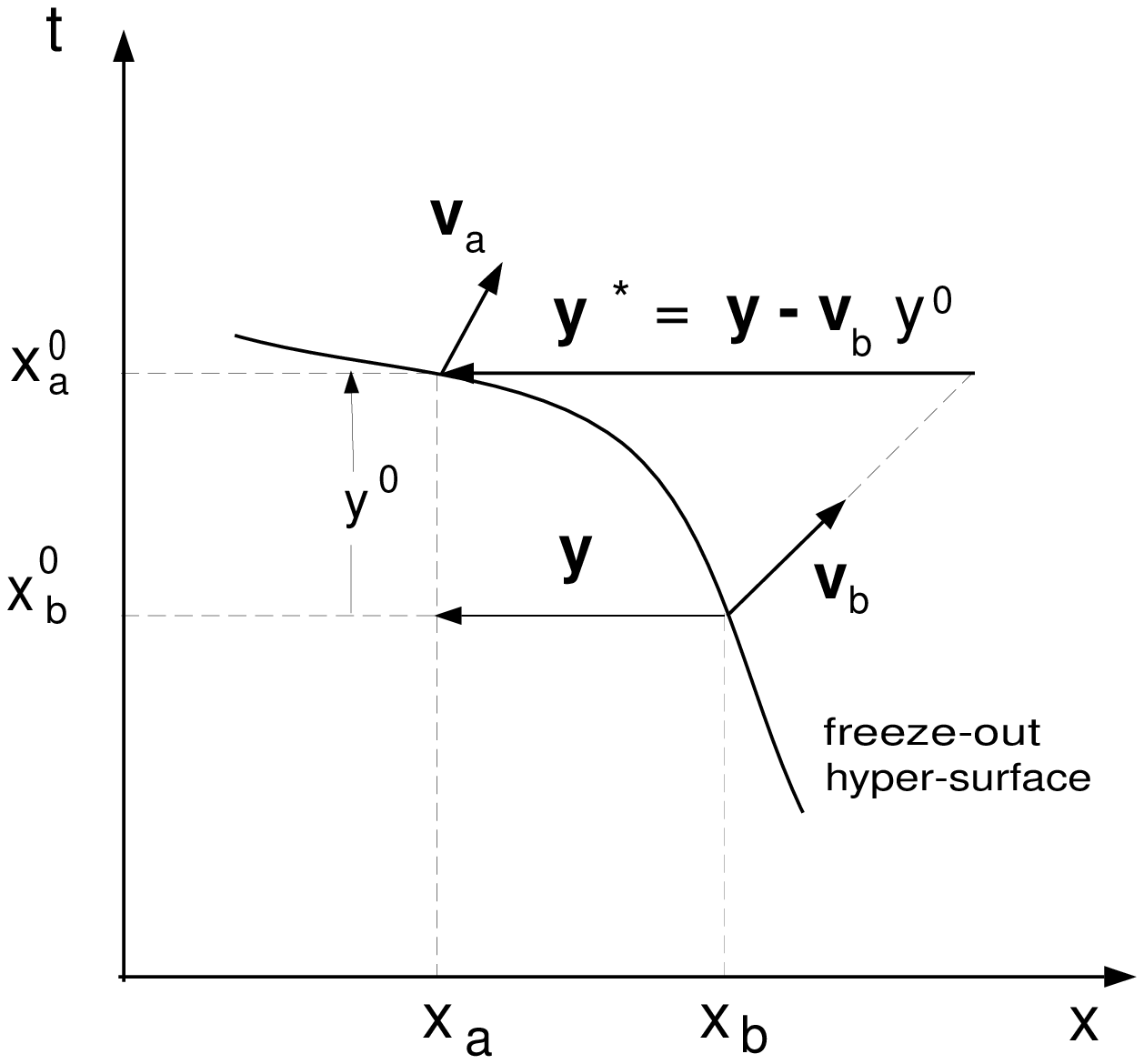}
\includegraphics{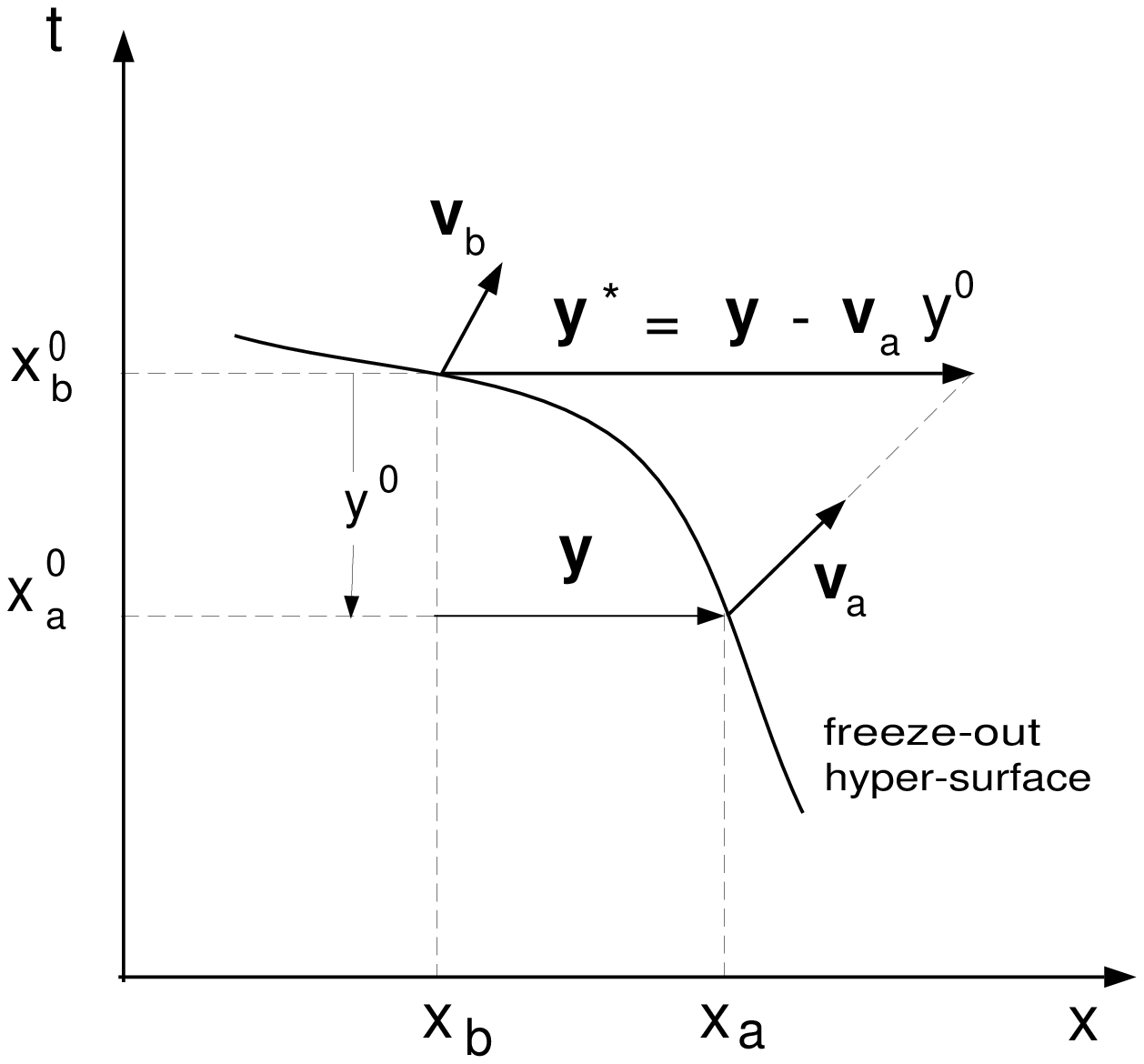}

%%%%%%%%%%%%% CAPTION%%%%%%%%%%%%%%%%%%%%%%%%%%%%%%%%%%%%%%%%%%%%
\begin{figure}
\caption{Graphical illustration of the coordinates at which the
FSI distorted waves $\phi$ in (\ref{50+6}) are evaluated. Upper
panel: the case $y^0 > 0$ (i.e. $x^0_a > x^0_b$). Lower panel: the case
$y^0 < 0$ (i.e. $x^0_a < x^0_b$).
}
\label{F1}
\end{figure}
%%%%%%%%%%%%%%%%%%%%%%%%%%%%%%%%%%%%%%%%%%%%%%%%%%%%%%%%%%%%%%%%%%%

%\begin{figure}[h]\epsfxsize=10cm 
%\centerline{\epsfbox{prlfig1.ps}}
%\vskip -6.5cm
%
%\vskip -70pt
%\begin{minipage}[t]{10cm}
%\noindent \bf Fig.1.  \rm
%The YKP radii $R_\perp$, $R_\parallel$, and $R_0$, for pairs with 
%$Y=0$ as functions of $M_\perp = (m^2+ K_\perp^2)^{1/2}$. 
%Dashed lines: pion correlations ($m=m_\pi$); solid lines: kaon 
%correlations ($m=m_K$). Left column: no transverse expansion of 
%source; right column: transverse expansion according to (\ref{27}) 
%with $\eta_f=0.6$. For more discussion see text.  
%\end{minipage}
%\vskip 4truemm

\end{document}